\documentclass[
  aps,
  prl,
  reprint,
  superscriptaddress,
  longbibliography,
  nofootinbib
]{revtex4-2}

\usepackage[T1]{fontenc}
\usepackage[utf8]{inputenc}
\usepackage[english]{babel}
\usepackage{amsmath,amssymb,bm}
\usepackage{graphicx}
\usepackage{xcolor}
\usepackage{booktabs}
\usepackage{array}
\usepackage{hyperref}
\usepackage{microtype}

\hypersetup{
  colorlinks=true,
  linkcolor=blue,
  citecolor=blue,
  urlcolor=blue
}

\newcommand{\AFMo}{\ensuremath{\mathrm{AFM}o}}

\newcommand{\LDSC}{\ensuremath{\mathrm{LDSC}}}

\newcommand{\funit}{\mathrm{f.u.}}
\newcommand{\mub}{\mu_{\mathrm B}}
\newcommand{\kb}{k_{\mathrm B}}

\begin{document}

\title{Competition between local magnetic disorder and altermagnetism in doped FeSb$_2$} 

\author{Enrico Di Lucente}
\author{Michele Simoncelli}
\affiliation{Department of Applied Physics and Applied Mathematics, Columbia University, New York 10027, USA}

\begin{abstract}
Recent experimental reports suggest that the narrow-gap nonmagnetic semiconductor FeSb$_2$ can be 
transformed into an altermagnetic metal through Co doping (Co$_{0.15}$Fe$_{0.85}$Sb$_2$), or into a magnetically disordered or short-range-ordered state through Cr doping (Cr$_{0.15}$Fe$_{0.85}$Sb$_2$).
Here we explore the energy landscape and magnetic states of these doped systems from first principles, relying on Hubbard-augmented density-functional theory (DFT+$U$) combined with the \texttt{Romeo} ground-state search algorithm. 
Within the established virtual-crystal approximation (VCA), we show that \texttt{Romeo} finds several non-trivial magnetic states, which inform targeted explicit simulations of doping in supercells.  We rely on these findings to discuss strengths and limitations of the VCA-\texttt{Romeo} approach versus the explicit-doping supercell approach, and how they can be used in synergy.
Overall, our simulations suggest that the ground state of the Cr-doped system is a Locally Disordered Spin-Compensated (\LDSC) configuration, formally compatible with Néel's L-type fully compensated ferrimagnetism, whereas the ground state of the Co-doped system is found to be altermagnetic (\AFMo). 
This work shows how approximate and explicit simulations of magnetic alloys can be mutually informative, and establishes a protocol for studying candidate metallic altermagnets.
\end{abstract}

\maketitle 

Antiferromagnetic spintronics aims to combine ultrafast dynamics and negligible stray fields with efficient electrical and optical readout \cite{smejkal2018topological,smejkal2022hallreview}. Altermagnets realize this combination as a third symmetry class of collinear magnetism: compensated opposite-spin sublattices are related by rotations, mirrors, screws, or glides rather than by translation or inversion. Altermagnetic signatures include momentum-dependent nonrelativistic spin splitting without net magnetization \cite{mazin2022editorial,smejkal2022phase,smejkal2022emerging,krempasky2024lifting,lee2024broken,takegami2025circular,reimers2024direct,yang2025mapping,santhosh2025thinfilms,zhou2025manipulation,rai2025polarity}
and chiral-magnon splitting~\cite{liu2024chiral,jost2025chiral,liu2026switchable,li2025xmcd,zhang2025crsbmagnon}. 

FeSb$_2$ has been predicted to offer a doping-controlled route to altermagnetic physics. Specifically, DFT with a phenomenological Hubbard correction and a VCA-based description of doping predicts that FeSb$_2$, a correlated narrow-gap semiconductor without static local-moment order \cite{petrovic2005kondo,zaliznyak2011absence,jie2012electronic,xu2020metallic}, becomes a metallic altermagnet upon weak Co or Cr doping \cite{mazin2021prediction}.
Following these predictions, experiments investigated Co$_{0.15}$Fe$_{0.85}$Sb$_2$~\cite{roy2026narrow} and Cr$_{0.15}$Fe$_{0.85}$Sb$_2$~\cite{shawon2026evidence}.
Experiments on Fe$_{1-x}$Co$_x$Sb$_2$ show that metallicity emerges near $x\simeq0.10$ without ferromagnetism; the 
$x=0.14(1)$ crystal is metallic, exhibits no magnetic hysteresis, and is nearly compensated (with a spontaneous magnetic moment below $3\times10^{-3}\,\mub/\funit$). 
Comparison of the experimental optical conductivity with VCA-based predictions for \AFMo, nonmagnetic, and conventional antiferromagnetic states showed the best agreement for \AFMo, leading to the conclusion that Co$_{0.15}$Fe$_{0.85}$Sb$_2$ may be altermagnetic.
These observations are consistent with earlier experiments \cite{hu2006anisotropy,hu2007weak}, which reported Fe$_{1-x}$Co$_x$Sb$_2$ to become metallic at $x\gtrsim 0.1$ and weak ferromagnetism to occur for approximately $0.20\le x\le0.45$.
In Fe$_{1-x}$Cr$_x$Sb$_2$, substitution near $x\simeq0.09$--$0.10$ creates Curie--Weiss moments without detectable order \cite{hu2007anisotropy}; at $x\simeq0.15$, broad local fields, the absence of coherent zero-field $\mu$SR oscillations and magnetic Bragg peaks, and a nearly compensated time-reversal-breaking response imply short-range or disordered magnetism rather than long-range \AFMo\ order \cite{shawon2026evidence}. Canted antiferromagnetism appears near $x\simeq0.25$ and persists toward $x\simeq0.45$, while antiferromagnetism develops at larger Cr content and culminates in the $T_N\simeq275$~K antiferromagnetic semiconductor CrSb$_2$ \cite{hu2007anisotropy}. 
 
Overall, these experimental reports show that many magnetic states compete in these systems and that VCA-based calculations, which effectively describe alloy doping in a mean-field way, are not always consistent with experimental findings. More precisely, the VCA describes Fe and the dopant using a compositionally averaged potential; it preserves the primitive-cell symmetry but cannot distinguish local Fe, Co, and Cr environments. 

The failure of the VCA to capture local inhomogeneities has recently been discussed in the literature \cite{smolyanyuk2025origin}. Specifically, in Cr-doped RuO$_2$, the VCA favors itinerant \AFMo\ over FM at 20\% Cr, whereas explicit simulations find FM lowest among the tested configurations, with Cr-dominated magnetic clusters and Ru moments that collapse away from Cr-rich regions, motivating a reinterpretation \cite{smolyanyuk2025origin} of the experimental results in Ref.~\cite{wang2023emergent}. That comparison considered conventional \AFMo\ and FM VCA states, while the explicit supercells were initialized as \AFMo, FM, or with only the Cr moments antiparallel. 

Here we advance simulations of magnetic alloys by introducing a workflow to investigate how alloy chemistry affects the magnetic states, simultaneously leveraging the following state-of-the-art computational techniques: (i) We combine the established VCA with the Robust Occupation Matrix Energy Optimization (\texttt{Romeo}) algorithm \cite{ponet2024landscape} to explore the magnetic energy landscape in full generality, without preselecting a magnetic class. Specifically, we show that a given VCA cell can sustain several self-consistent magnetic states, including ferrimagnetic (FiM) minima that are not evident a priori. (ii) We rely on the VCA states to initialize analogous non-trivial magnetic states in 120-atom $4\times1\times5$ supercells where alloy disorder is explicitly described and global \AFMo\ order is possible at the experimental Co~\cite{roy2026narrow} or Cr~\cite{shawon2026evidence} concentration $x=0.15$. Importantly, our framework relies on Hubbard-augmented DFT \cite{cococcioni2005linear,timrov2018hubbard}, with the $U$ parameter computed from first principles \cite{timrov2021self,timrov2022hp}, to mitigate the self-interaction error of local and semilocal DFT for localized Fe $3d$ electrons.

\begin{figure}[t]
\centering
\includegraphics[width=\columnwidth]{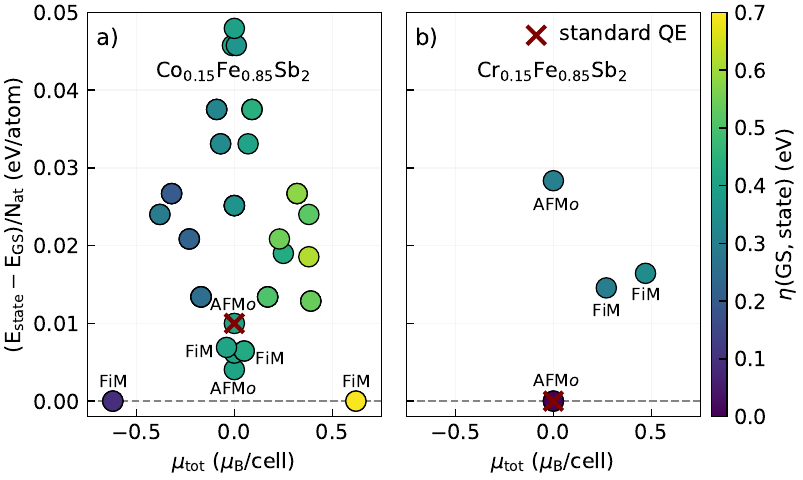}
\caption{\textbf{DFT+$\bm{U}$+VCA magnetic-energy landscapes.} Self-consistent solutions obtained with \texttt{Romeo} \cite{ponet2024landscape} for (a) Co$_{0.15}$Fe$_{0.85}$Sb$_2$ and (b) Cr$_{0.15}$Fe$_{0.85}$Sb$_2$ at $U=5$~eV computed from first principles~\cite{timrov2021self}. Each point represents a local minimum of the unconstrained DFT+$U$+VCA functional. The vertical axis gives the energy per atom relative to the lowest-energy solution of the corresponding composition, while the horizontal axis gives the total magnetic moment per six-atom cell. The color encodes the occupied-band distance $\eta$ from the corresponding ground state (see text). Red crosses denote conventional single-start \texttt{Quantum ESPRESSO} solutions; in panel (a), the single-start \AFMo\ solution is the fifth excited state rather than the lowest \AFMo\ minimum. 
}
\label{fig:romeo}
\end{figure}

\textit{Simulations of doping with the VCA-\texttt{Romeo} approach.} 
In Fig.~\ref{fig:romeo} we analyze the magnetic energy landscape obtained with the VCA-\texttt{Romeo} approach at the value $U\simeq5$~eV computed from first principles (Supplemental Fig.~\ref{fig:plot_Hubbard_U_vs_iterations}). Each point is a different self-consistent occupation-matrix solution, colored by its occupied-band distance $\eta$ from the corresponding ground state, defined as the occupation-weighted root-mean-square difference between the Kohn--Sham eigenvalues over the full Brillouin zone, minimized with respect to a rigid energy shift \cite{prandini2018precision}. 
Within the VCA, the Co-doped ground state is weakly ferrimagnetic (FiM), with approximately $0.3\,\mub/\funit$, whereas the Cr-doped ground state is \AFMo. 
For Co doping at $U=5$~eV, a conventional single-start calculation fails to recover the lowest \AFMo\ minimum and instead converges to an \AFMo\ solution corresponding to the fifth excited state in the energy landscape found by \texttt{Romeo}.
To assess the sensitivity of the results to the Hubbard parameter, we investigate the effect of varying $U$ from 0 to $5$~eV and find that the VCA ground state switches nonmonotonically between \AFMo\ and FiM minima for both alloys (Supplemental Figs.~\ref{fig:Co_romeo_Uscan} and \ref{fig:Cr_romeo_Uscan}). 

X-ray absorption spectroscopy reveals that FeSb$_2$ exhibits a predominantly low-spin magnetic character \cite{li2024spectroscopic}, while the Co-doped sample at $x\simeq0.15$ is nearly magnetically compensated \cite{roy2026narrow}. In our Co-doped VCA calculations, low-spin FM configurations are never ground states. At the $U$ value computed self-consistently for the \AFMo\ state, \texttt{Romeo} finds a lower-energy FM solution only after a pronounced redistribution of the Fe-$3d$ occupations toward a high-spin state. This FM solution is inconsistent with experiment and emerges through the known DFT+$U$ bias toward high-spin solutions \cite{mariano2020biased,mariano2021improved,di2026spin}. We therefore discard this FM state. Thus, a single VCA cell does not define a unique self-consistent magnetic state, and to check the physical reliability of its lowest-energy minimum we perform explicit-doping simulations in supercells.

\textit{Explicit simulations of doping in supercells.---} FeSb$_2$ has space group $Pnnm$; at $x=0.15$ the minimal cell containing an integer number of dopants and supporting an exact global \AFMo\ operation is a $4\times1\times5$ replication, Fe$_{34}M_6$Sb$_{80}$ ($M=$ Co or Cr), with 120 atoms. Requiring one nontrivial parent operation to exchange the complete up and down sublattices while preserving Fe, $M$, and Sb reduces $\binom{20}{3}^{2}$ compensated assignments to 342 inequivalent \AFMo\ classes, independently verified with \texttt{AMCheck} \cite{amcheck1,amcheck2}. Under the same equivalence, the complete compensated space contains $8\,334$ classes, of which $7\,992$ are non-\AFMo; we therefore characterize the smaller ordered manifold first and access physically motivated \LDSC\ motifs along shortest atomic-swap paths rather than relaxing every non-\AFMo\ class (Sec.~\ref{sec:ldsc_supp}).

\begin{figure}[t]
\centering
\includegraphics[width=\columnwidth]{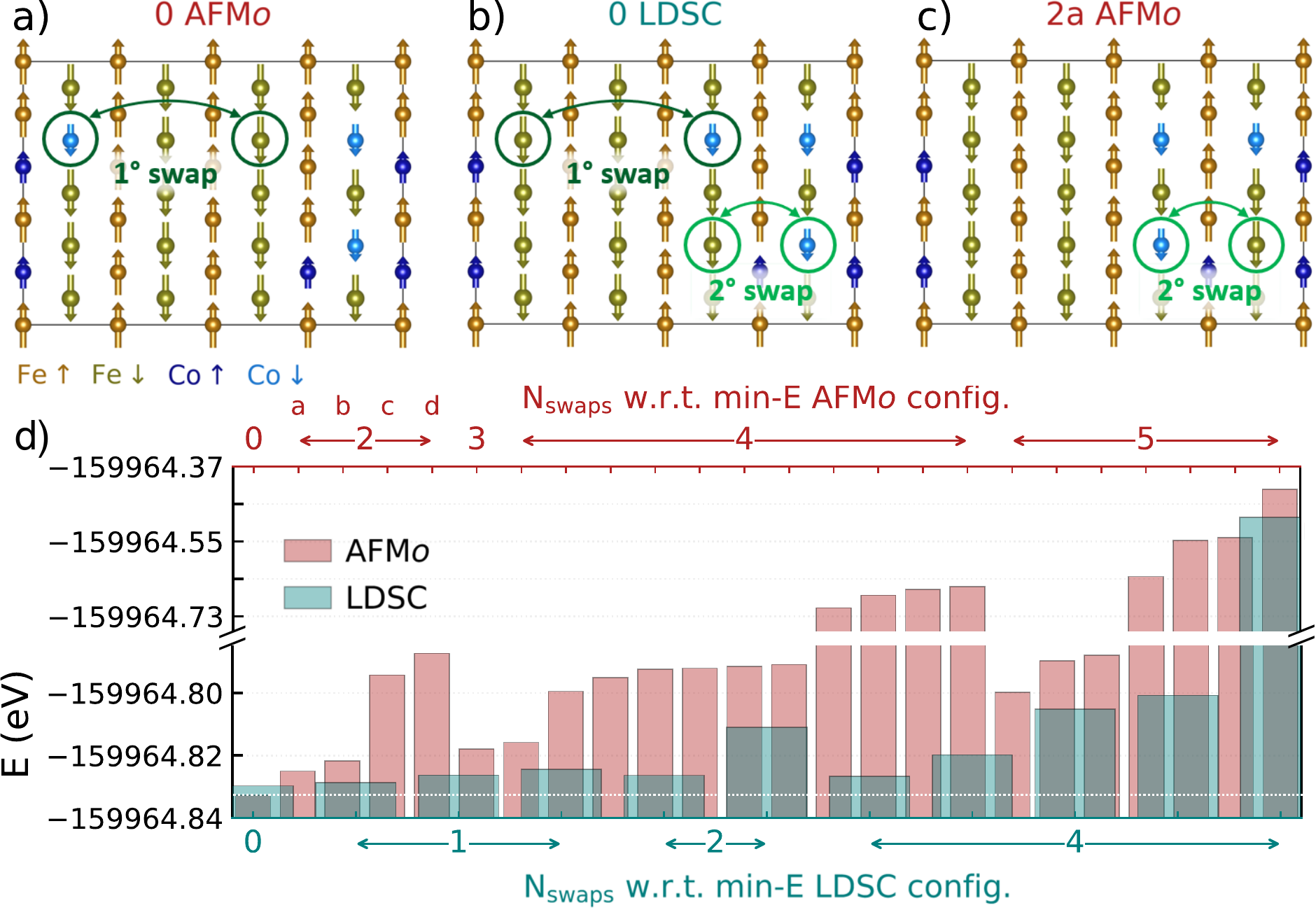}
\caption{\textbf{Configurations and energy landscape for \textbf{Co$_{0.15}$Fe$_{0.85}$Sb$_2$}.} (a) Minimum-energy \AFMo\ configuration, denoted 0~\AFMo. (b) Minimum-energy \LDSC\ configuration, denoted 0~\LDSC, reached from (a) by one same-spin Fe--Co occupation swap. (c) The displayed 2a~\AFMo\ configuration lies two swaps from the \AFMo\ minimum and one swap from (b). Green circles and arrows mark the exchanged sites; Fe$\uparrow$, Fe$\downarrow$, Co$\uparrow$, and Co$\downarrow$ are shown in gold, olive, dark blue, and light blue, respectively. (d) DFT+$U$ total energies of the converged \AFMo\ (red) and \LDSC\ (blue) configurations, ordered by their minimum swap distance from the corresponding family minimum.}
\label{fig:Co_landscape}
\end{figure}

\begin{figure}[t]
\centering
\includegraphics[width=\columnwidth]{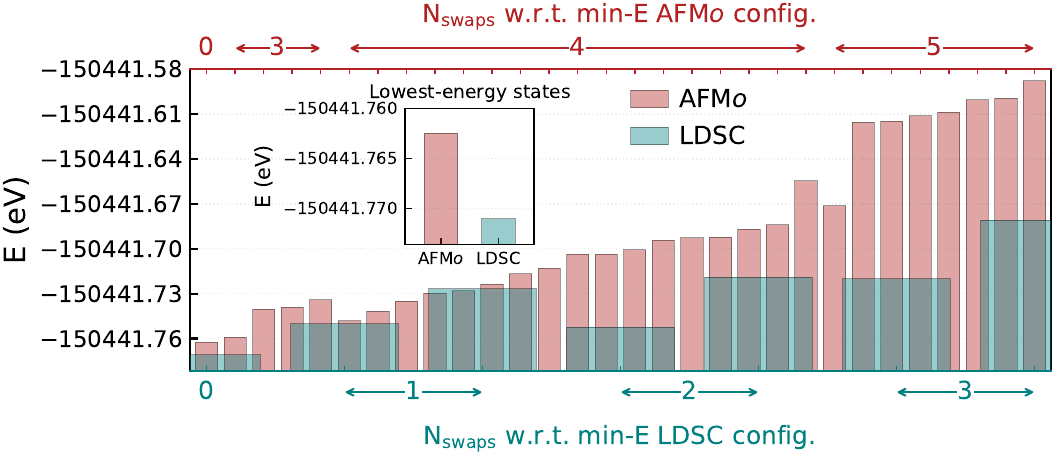}
\caption{\textbf{Energy landscape for \textbf{Cr$_{0.15}$Fe$_{0.85}$Sb$_2$}.} Absolute DFT+$U$ total energies of converged \AFMo\ (red) and \LDSC\ (blue) configurations, ordered by the minimum number of atomic swaps from their respective family energy minima. The upper and lower annotations give the swap distances, and the inset compares the two lowest-energy states. The \LDSC\ minimum lies below the lowest sampled \AFMo\ state. }
\label{fig:Cr_landscape}
\end{figure}

Figures~\ref{fig:Co_landscape} and \ref{fig:Cr_landscape} show the converged \AFMo\ and \LDSC\ configurations. Within each family, states are ordered by the exact minimum number of atomic swaps from that family's lowest-energy state; configurations at equal distance are sorted by energy. We found that the ground state (the lowest-energy state identified within our explicit-alloy search) is \AFMo\ for Co doping and \LDSC\ for Cr doping, with energy separations of $\simeq3$ and $\simeq8.5$~meV per 120-atom cell, respectively. In Fig.~\ref{fig:Co_landscape}(a--c), one swap connects the minimum-energy \AFMo\ and \LDSC\ motifs, and a second reaches one displayed \AFMo\ configuration at swap distance two. The uncompensated FiM minimum found by VCA-\texttt{Romeo}, produced by unequal Fe and dopant moment magnitudes, motivates explicit-alloy initializations with uncompensated sublattices. No such state survives as a self-consistent explicit-alloy phase: FiM-derived initializations either reclassify to a different magnetic class or remain well above the \AFMo/\LDSC\ manifold (see Supplemental Sec.~\ref{sec:magnetic_initializations}). Thus, the VCA landscape guides targeted supercell sampling, while the explicit cell tests which magnetic states survive when explicitly accounting for doping.

The \LDSC\ motif is the first non-\AFMo\ state on a shortest allowed atomic-site swap path between low-energy \AFMo\ decorations [Fig.~\ref{fig:Co_landscape}(a--c)]. The swap preserves the composition, the numbers of spin-up and spin-down labels, and most local antiparallel Fe--Fe and Fe--dopant environments, while removing all symmetry operations that map the entire spin-up sublattice onto the spin-down one, thus precluding altermagnetism. \LDSC\ is therefore locally symmetry-disordered. At $T=0$~K, its inequivalent antiparallel sublattices and numerically vanishing total moment make it formally compatible with Néel's L-type, or fully compensated ferrimagnetic, class \cite{neel1948proprietes,neel1984magnetic,fecher2026compensated}; we note, in passing, that strict L-type thermomagnetic behavior would require finite-temperature sublattice magnetizations. A nonperiodic distribution of these low-cost motifs can disrupt long-range \AFMo\ coherence while preserving local antiparallel correlations, making \LDSC\ the closest static DFT proxy examined here for the short-range-disordered Cr state and therefore qualitatively consistent with the experimentally observed nearly spin-compensated magnetism \cite{shawon2026evidence}. The periodic supercell used is an ordered computational snapshot and does not determine a magnetic correlation length.

These explicit calculations also rule out the mechanism in which the host remains essentially nonmagnetic and magnetism resides on the dopant, discussed for Cr-doped RuO$_2$~\cite{smolyanyuk2025origin}. Initializations with Fe constrained or initialized nonmagnetically do not produce a stable or low-energy solution: Fe moments re-emerge or the state remains far above the \AFMo/\LDSC\ manifold. Thus, within the DFT+$U$ manifold, low-energy Fe$_{0.85}M_{0.15}$Sb$_2$ maintains an intrinsically magnetic Fe network, in contrast to the weakly polarizable Ru host of Cr-doped RuO$_2$~\cite{smolyanyuk2025origin}.

\begin{figure}[t]
\centering
\includegraphics[width=\columnwidth]{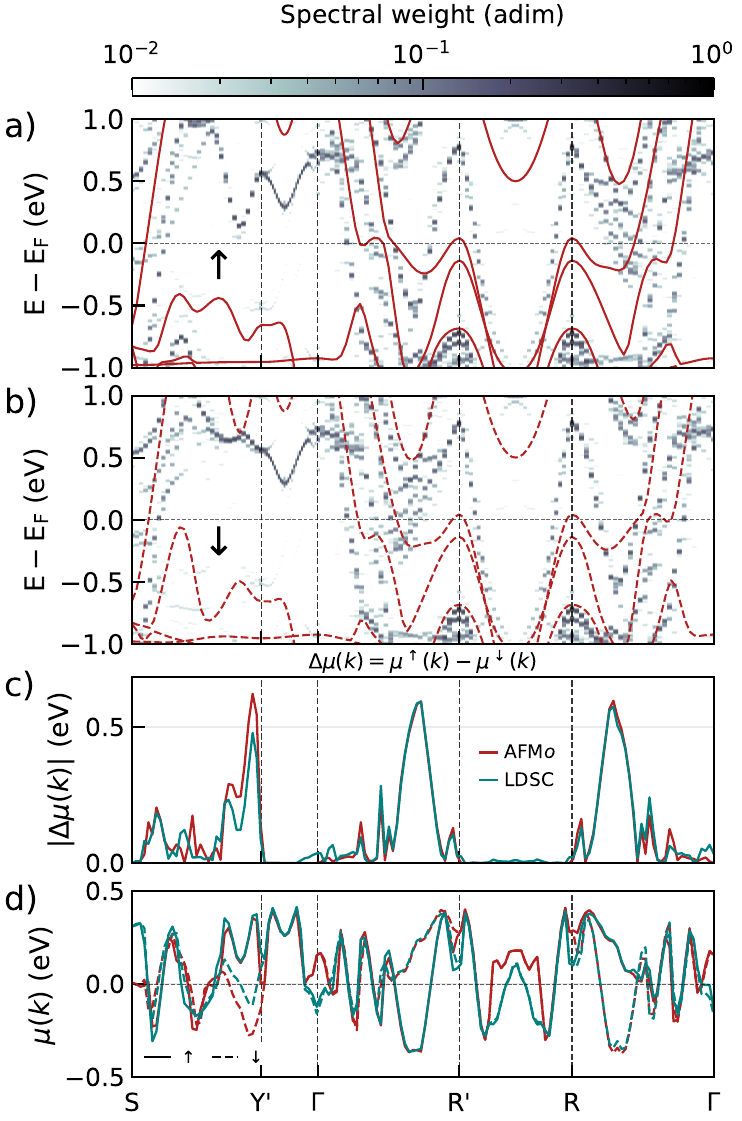}
\caption{\textbf{Electronic structure for \textbf{Co$_{0.15}$Fe$_{0.85}$Sb$_2$}.} (a,b) Spin-up and spin-down unfolded spectral weights of the lowest-energy \AFMo\ supercell; the overlaid red curves are the $U=5$~eV VCA bands of the lowest-energy \AFMo\ state in Fig.~\ref{fig:romeo}(a). (c) Magnitude of the centroid spin splitting $|\Delta\mu(k)|=|\mu_\uparrow(k)-\mu_\downarrow(k)|$ for \AFMo\ (red) and \LDSC\ (blue). (d) Spin-resolved centroids defined in Eq.~\eqref{eq:centroid_main}: \AFMo\ is red, \LDSC\ is blue, spin up is solid, and spin down is dashed. The centroids are evaluated over $-0.5\le E-E_F\le0.5$~eV and compare the energy center of the unfolded near-Fermi spectral weight.}
\label{fig:Co_bands}
\end{figure}

\begin{figure}[t]
\centering
\includegraphics[width=\columnwidth]{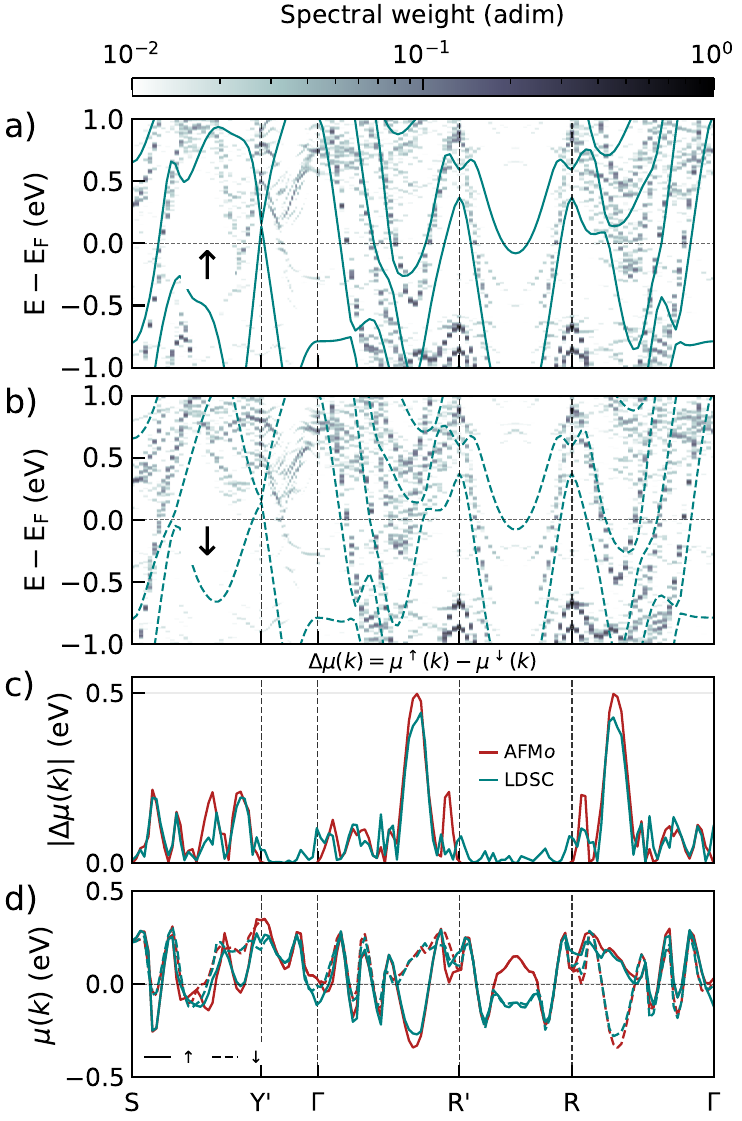}
\caption{\textbf{Electronic structure for \textbf{Cr$_{0.15}$Fe$_{0.85}$Sb$_2$}.} (a,b) Spin-up and spin-down unfolded spectral weights of the lowest-energy \LDSC\ supercell; the overlaid blue curves are the $U=5$~eV VCA bands of the lowest-energy \AFMo\ state in Fig.~\ref{fig:romeo}(b) (the \LDSC\ state cannot be realized within the unit-cell VCA description). (c) $|\Delta\mu(k)|$ for the lowest sampled \AFMo\ state (red) and the \LDSC\ ground state (blue). (d) The corresponding spin-resolved centroids, with solid and dashed lines denoting spin up and down. The same $[-0.5,0.5]$~eV window, broadening, momentum path, and Fermi-level convention are used for both states.}
\label{fig:Cr_bands}
\end{figure}

Figures~\ref{fig:Co_bands} and \ref{fig:Cr_bands} compare the spin-resolved unfolded spectra~\cite{mondal2026banduppy,medeiros2014bandunfolding,medeiros2015spinor,iraola2022irrep} of the two competing states (see Supplemental Sec.~\ref{sec:unfolding_supp} for details). To compare broadened alloy spectra, we use the first moment of the near-Fermi spectral weight:
\begin{equation}
\mu_\sigma(k)=\frac{\int_{-0.5\,\mathrm{eV}}^{0.5\,\mathrm{eV}} E A_\sigma(k,E)\,dE}{\int_{-0.5\,\mathrm{eV}}^{0.5\,\mathrm{eV}} A_\sigma(k,E)\,dE},
\label{eq:centroid_main}
\end{equation}
where energies are measured from $E_F$ and $A_\sigma(k,E)$ is the spin-resolved unfolded spectral function. The centroid tracks the energy center of the spectral weight in a fixed window, so $|\Delta\mu(k)|=|\mu_\uparrow(k)-\mu_\downarrow(k)|$ provides a measure of spin splitting even when disorder blurs or fragments the bands.

For Co$_{0.15}$Fe$_{0.85}$Sb$_2$, Figs.~\ref{fig:Co_bands}(a,b) show the spin-up and spin-down spectra of the \AFMo\ ground state. Panel (c) compares $|\Delta\mu(k)|$ for \AFMo\ and \LDSC, and panel (d) shows the four underlying centroids: \AFMo\ in red, \LDSC\ in blue, with solid and dashed lines for spin up and down. The \LDSC\ swap leaves the near-Fermi dispersion mostly unchanged but modifies the centroid splitting at selected wavevectors, with a pronounced change in $|\Delta\mu(\mathbf{k})|$ near $\mathrm{Y}'$ along the $\mathrm{S}$--$\mathrm{Y}'$ path, while removing the altermagnetic correspondence between symmetry-related opposite-spin spectra.

For Cr$_{0.15}$Fe$_{0.85}$Sb$_2$, Figs.~\ref{fig:Cr_bands}(a,b) instead show the spin-resolved spectra of the \LDSC\ ground state. Panels (c,d) compare its centroid splitting and spin-resolved centroids with those of the lowest sampled \AFMo\ state. Analogously to the Co-doped case, the two configurations maintain similar broad dispersive features but differ slightly, mainly along the $\Gamma$--$\mathrm R'$ and $\mathrm R$--$\Gamma$ directions of the path. These comparisons separate symmetry-defined \AFMo\ splitting from generic spin-resolved reconstruction in a compensated alloy cell.

To probe chemical and magnetic disorder, we also consider the \AFMo\ and \LDSC\ classes together with three special quasirandom structure (SQS) reference classes representing chemical, magnetic, and combined disorder.
To test whether configurational multiplicity can offset energy differences among the sampled states, we characterize each disordered-cell class $i$ by a multiplicity $\Omega_i$ and a DFT+$U$ energy $E_i$. 
The class free energy can be approximately defined as $A_i(T)=E_i-\kb T\ln\Omega_i$ (the approximation lies in neglecting vibrational, electronic, spin-wave, and dynamical magnetic entropies, see Supplemental Sec.~\ref{sec:free_energy_supp}).
For Cr doping, entropy is not needed to favor \LDSC\ because it already has the lowest energy at $T=0$~K. For Co doping, the larger \LDSC\ multiplicity lowers $A_i(T)$ below the ordered \AFMo\ class near $25$~K.
The crossing near $25$~K only means that, in this approximate expression, the larger number of \LDSC\ realizations compensates the higher energy of the \LDSC\ class. It does not predict a phase transition, because the Fe and Co atoms cannot exchange sites at such low temperatures; their arrangement was fixed during synthesis.
The three SQS reference models \cite{zunger1990sqs,di2026spin} lie much higher in energy, and their greater multiplicities offset this energy cost only at substantially higher temperatures (see Supplemental Fig.~\ref{fig:free_energy_supp}).

We conclude by noting that the 120-atom periodic cell resolves the local doping chemistry while representing only one member of the disorder ensemble.
Our simulations test whether selected explicit local rearrangements destabilize a symmetry-defined \AFMo\ reference. A formal chemistry-first approach would apply \texttt{enumlib} \cite{hart2008algorithm,hart2009generating,hart2012generating} to every diagonal and non-diagonal index-20 supercell and then explore the magnetic states of each decoration. Such an unrestricted chemistry-first search was applied to Cr-doped RuO$_2$ at $x=0.2$ using standard DFT \cite{smolyanyuk2025origin}, leading to 52 inequivalent supercells, each tested with three magnetic initializations. However, this framework would be prohibitive here because the smaller doping concentration $x=0.15$ requires supercells containing at least 120 atoms: even after reduction by translations and parent-space-group symmetries and the removal of superperiodic structures, $4,109$ symmetry-inequivalent chemical decorations would remain to be studied. A complete landscape-to-landscape comparison would additionally apply \texttt{Romeo} to every explicit decoration. Since this approach is computationally prohibitive with the resources at our disposal, we employ the VCA-\texttt{Romeo} approach to identify nontrivial magnetic states that are not evident from conventional \AFMo/FM calculations, and use these to selectively test analogous magnetic states in explicit-alloy supercells.

\textit{Conclusions.---} In summary, we have shown that combining VCA simulations within Hubbard-augmented DFT (with the Hubbard parameter computed from first principles \cite{timrov2021self}) with the \texttt{Romeo} \cite{ponet2024landscape} search algorithm allows us to search for magnetic states beyond conventional \AFMo/FM initializations, revealing nontrivial (e.g., uncompensated FiM) minima that can compete with altermagnetic order. These nontrivial VCA states provide informed starting points for explicit supercell calculations, which test whether analogous states survive when the local doping chemistry is fully taken into account. Here, we found VCA FiM states that motivate corresponding uncompensated supercell initializations. Calculations from these initializations either converge to a different magnetic class or to states that remain high in energy. For both dopants, the targeted supercell search yields a lowest-energy sampled magnetic class that differs from the VCA result.
For Cr$_{0.15}$Fe$_{0.85}$Sb$_2$, we show that Cr doping stabilizes a compensated motif with local symmetry disorder already at zero temperature, providing a first-principles microscopic picture consistent with the observed short-range or disordered state \cite{shawon2026evidence} and making long-range altermagnetism unlikely. For Co$_{0.15}$Fe$_{0.85}$Sb$_2$, we show that Co doping stabilizes \AFMo\ at zero temperature. This finding contrasts with the VCA FiM solution and is consistent with a recent experimental report \cite{roy2026narrow}, although the larger \LDSC\ multiplicity can offset its small energy penalty at finite temperature in the configurational analysis performed. 
Direct diffraction or local-probe measurements on Co-doped crystals, together with optical-response calculations, are the natural tests of our predictions. Spin-resolved magnon splitting may offer a complementary signature of altermagnetic or fully compensated ferrimagnetic order, while site-resolved magnetic exchange parameters could clarify whether magnetic frustration is the microscopic origin of the contrasting response to electron and hole doping. These complementary analyses will be discussed in a separate work \cite{moorhouse2026manuscript}. Overall, we have shown that combining the VCA-\texttt{Romeo} magnetic-state-search approach with targeted explicit simulations of alloy doping in supercells provides a tractable methodology to quantitatively investigate the magnetic landscape of candidate metallic altermagnets: VCA-\texttt{Romeo} reveals magnetic complexity and guides the supercell search, while explicit supercells resolve the local chemical environments absent from the VCA.

\clearpage

\begin{center}
 {\large\bfseries Supplemental Material for\\[3pt]
 ``Competition between local magnetic disorder and altermagnetism in doped FeSb$_2$''}
\end{center}

\setcounter{equation}{0}
\renewcommand{\theequation}{S\arabic{equation}}
\renewcommand{\theHequation}{S\arabic{equation}}
\setcounter{figure}{0}
\renewcommand{\thefigure}{S\arabic{figure}}
\renewcommand{\theHfigure}{S\arabic{figure}}
\setcounter{table}{0}
\renewcommand{\thetable}{S\arabic{table}}
\renewcommand{\theHtable}{S\arabic{table}}

\setcounter{secnumdepth}{2}
\setcounter{section}{0}

\renewcommand{\thesection}{S\Roman{section}}
\renewcommand{\thesubsection}{\thesection.\Alph{subsection}}

\makeatletter
\renewcommand{\p@subsection}{}
\makeatother

\renewcommand{\theHsection}{supp.\arabic{section}}
\renewcommand{\theHsubsection}{supp.\arabic{section}.\arabic{subsection}}

\section{Computational details}

Calculations were performed with plane-wave density-functional theory as implemented in \texttt{Quantum ESPRESSO} \cite{giannozzi2009quantum,giannozzi2017advanced}. We used the PBE functional, a rotationally invariant DFT+$U$ correction with ortho-atomic Hubbard projectors, and the same PBE pseudopotential set for all compared configurations. Primitive-cell and VCA calculations used a $10\times10\times20$ Monkhorst--Pack mesh; the $4\times1\times5$ supercells used $2\times8\times3$. Brillouin-zone integrations employed Marzari--Vanderbilt smearing of width $0.02$~Ry.

All pseudopotentials used in these calculations were taken from the PS Library (version 2.0.1). Specifically, we selected PBE scalar-relativistic ultrasoft pseudopotentials (USPPs). Spin--orbit coupling (SOC) was not included, as scalar-relativistic potentials are sufficient to observe nonrelativistic altermagnetic band splitting. Moreover, the effect of SOC was reported to induce shifts in energy differences of less than 0.2 meV \cite{mazin2021prediction}. This is an order of magnitude smaller than the energy differences obtained in this study, which determine the stability hierarchy of the various magnetic phases investigated. For the VCA generation, we used \texttt{Quantum ESPRESSO}'s \texttt{virtual\_v2.x} code, which requires all input pseudopotentials to be of exactly the same type and currently lacks support for projector-augmented-wave (PAW) pseudopotentials. The original elemental pseudopotentials were used for the explicit-supercell calculations.

The effective Hubbard $U$ parameter was obtained by linear-response DFPT \cite{cococcioni2005linear,timrov2018hubbard,timrov2021self,timrov2022hp}. Repeating DFPT for every 120-atom decoration is prohibitive, so the response was evaluated for the VCA reference and a common $U=5$~eV was used for Fe and the dopant in the final comparisons. The converged \AFMo\ values are approximately $5.1$~eV for Co doping and $5.3$~eV for Cr doping. VCA calculations were additionally repeated for integer $U$ values from $0$ to $5$~eV to expose the sensitivity of the magnetic landscape.

To reliably resolve small energy differences ($\sim3$~meV) between configurations while managing the high computational cost, the wave-function and charge-density cutoffs were set to 90 and 1080~Ry, respectively, with a self-consistency threshold of $1.0\times10^{-6}$~Ry. This setup yields a residual numerical error of approximately $0.01$~meV, two orders of magnitude smaller than the target energy scale. To validate this choice, the threshold was further tightened to $1.0\times10^{-8}$~Ry for the lowest-energy states, confirming that the energies were already converged.

\section{Virtual-crystal approximation}
\label{sec:vca_supp}

\subsection{DFPT Hubbard interaction and conventional magnetic branches}

\begin{figure}[t]
\centering
\includegraphics[width=\columnwidth]{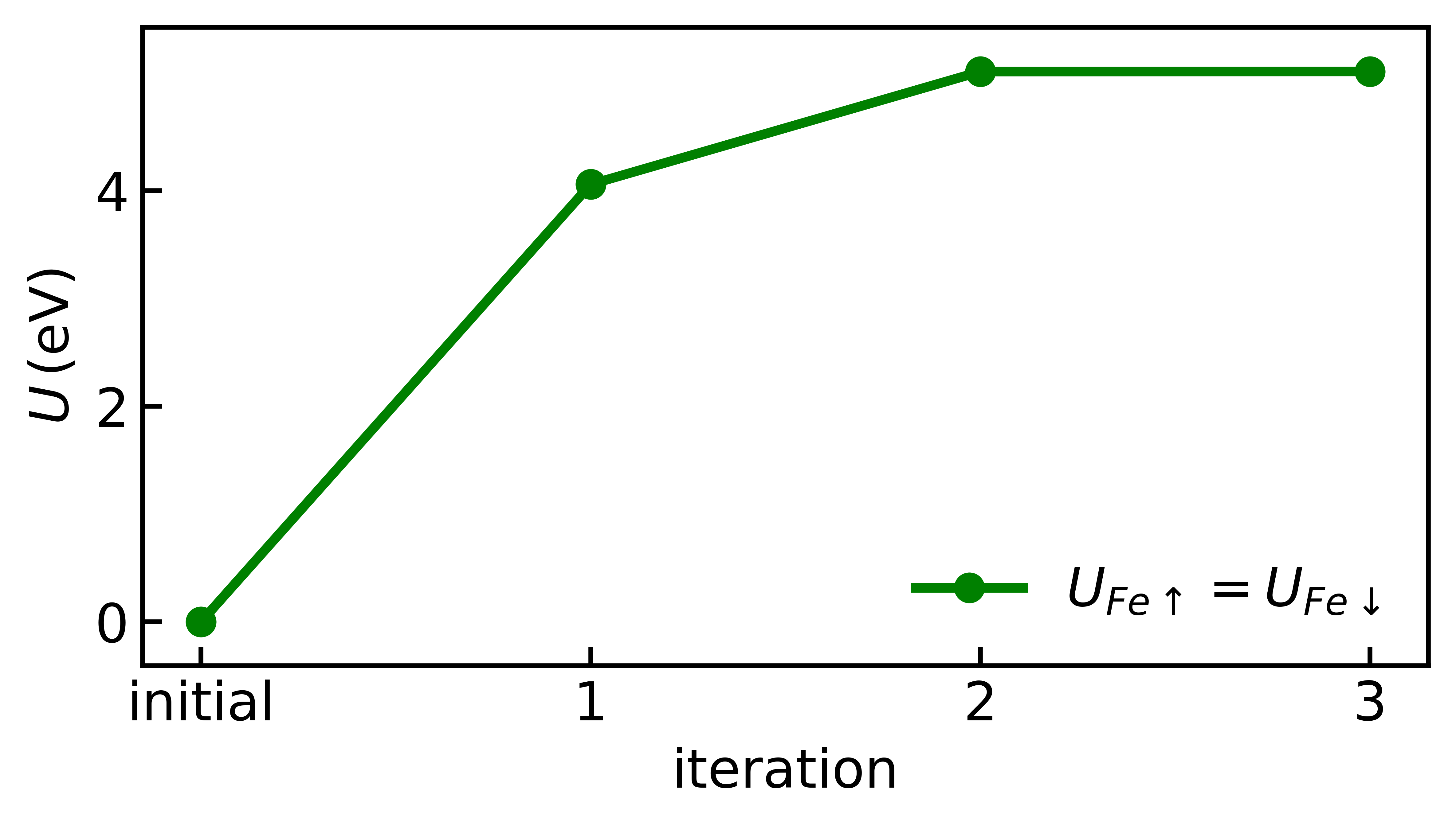}
\caption{\textbf{Iterative DFPT Hubbard interaction.} Convergence of the Hubbard $U$ parameter for the \AFMo\ configuration (see text for other magnetic orderings) of the Co-doped compound modeled with the VCA, yielding $U=5.1$~eV.}
\label{fig:plot_Hubbard_U_vs_iterations}
\end{figure}

Within VCA, the linear-response Hubbard parameter is an effective interaction of the compositionally averaged transition-metal site, not a species- or environment-resolved Fe/Co/Cr parameter. For Co-doped \AFMo\ VCA, the procedure converges near $5.1$~eV (see Fig.~\ref{fig:plot_Hubbard_U_vs_iterations}). Starting from the corresponding FiM solution yields inequivalent effective responses of approximately $4.81$ and $5.69$~eV on the two VCA Fe sites, due to their unequal magnetic moments. For a ferromagnetic initialization, the first response from $U=0$ is about $4.06$~eV; retaining the low-spin-like branch at this value produces a subsequent response near $5.93$~eV, which drives the solution toward the high-spin-like occupation pattern. We therefore use the common rounded value $U=5$~eV for controlled comparisons and treat the full $U$ scan, rather than any single branch, as the diagnostic of VCA robustness.

\begin{figure}[t]
\centering
\includegraphics[width=\columnwidth]{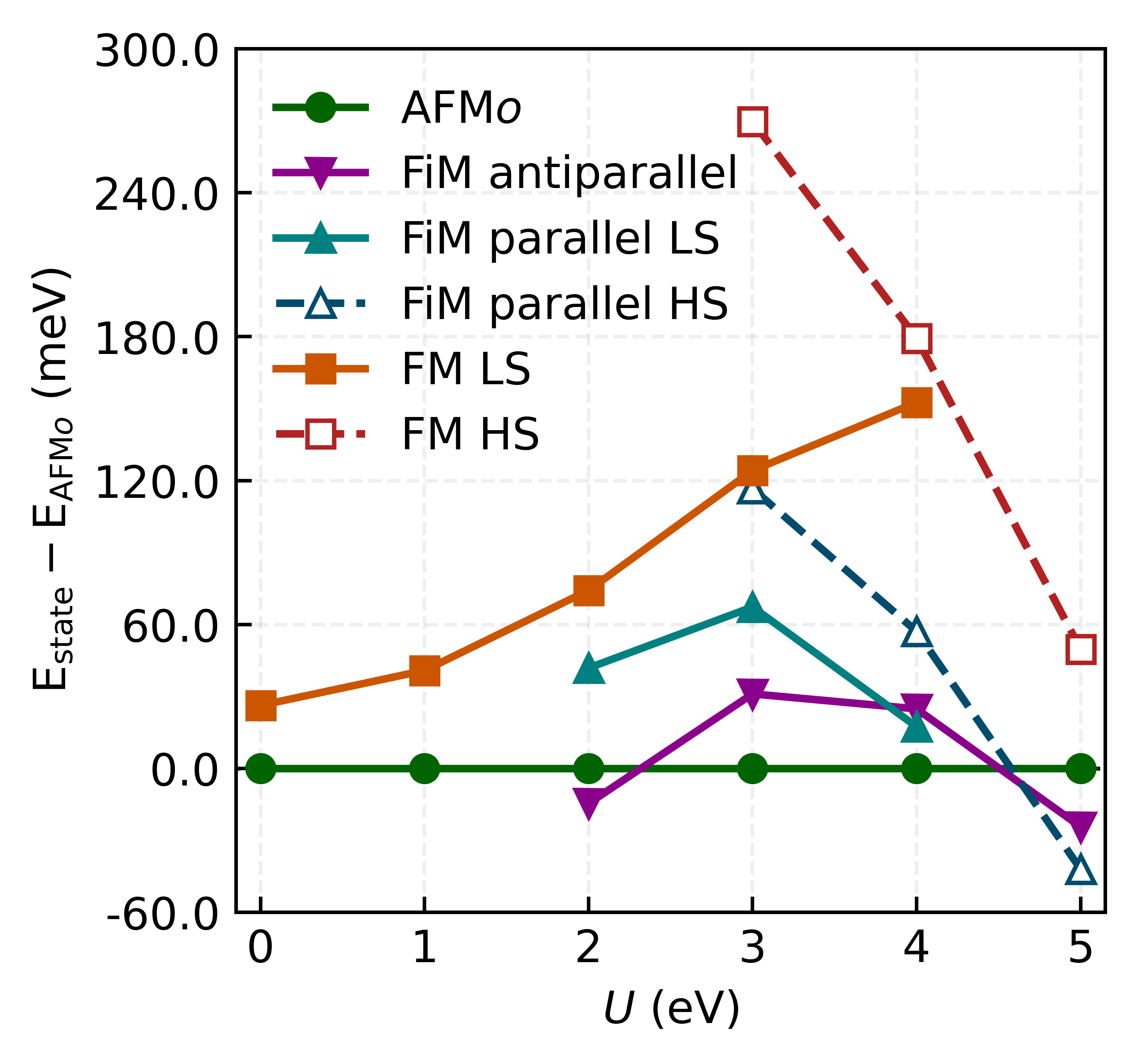}
\caption{\textbf{Conventional VCA magnetic states of the Co-doped compound versus $\bm{U}$.} Energies of \AFMo, antiparallel FiM, parallel FiM, and FM solutions relative to \AFMo. LS FM solutions are never lowest. HS-like parallel branches appear only after a large-$U$ redistribution of the Fe-$3d$ occupations.}
\label{fig:energy_vs_U}
\end{figure}

Figure~\ref{fig:energy_vs_U} compares conventional \AFMo, FiM, and FM branches relative to \AFMo\ at each $U$. ``Low spin'' (LS) and ``high spin'' (HS) are descriptive labels for a metal, not exact ionic quantum numbers. All LS ferromagnetic branches remain above the antiparallel \AFMo/FiM\ branches. The apparently favorable high-$U$ FM state emerges only after the Fe-$3d$ occupation rearrangement documented below and is inconsistent with the predominantly low-spin spectroscopic character of FeSb$_2$ \cite{li2024spectroscopic} and with the nearly compensated Co-doped experiment \cite{roy2026narrow}. This behavior is a recognized DFT+$U$ spin-state pathology: the Hubbard energy can penalize the more fractional occupations of low-spin, covalent states and systematically favor high spin \cite{mariano2020biased,mariano2021improved}; conversely, a high-spin-like response can yield a larger effective $U$, which further stabilizes integer occupations. An analogous self-reinforcing high-spin/$U$ feedback was found in FeSb$_3$ \cite{di2026spin}. We therefore regard the high-$U$ FM branch as a methodological warning rather than evidence for the experimental magnetic state.

\begin{figure*}[t]
\centering
\includegraphics[width=0.96\textwidth]{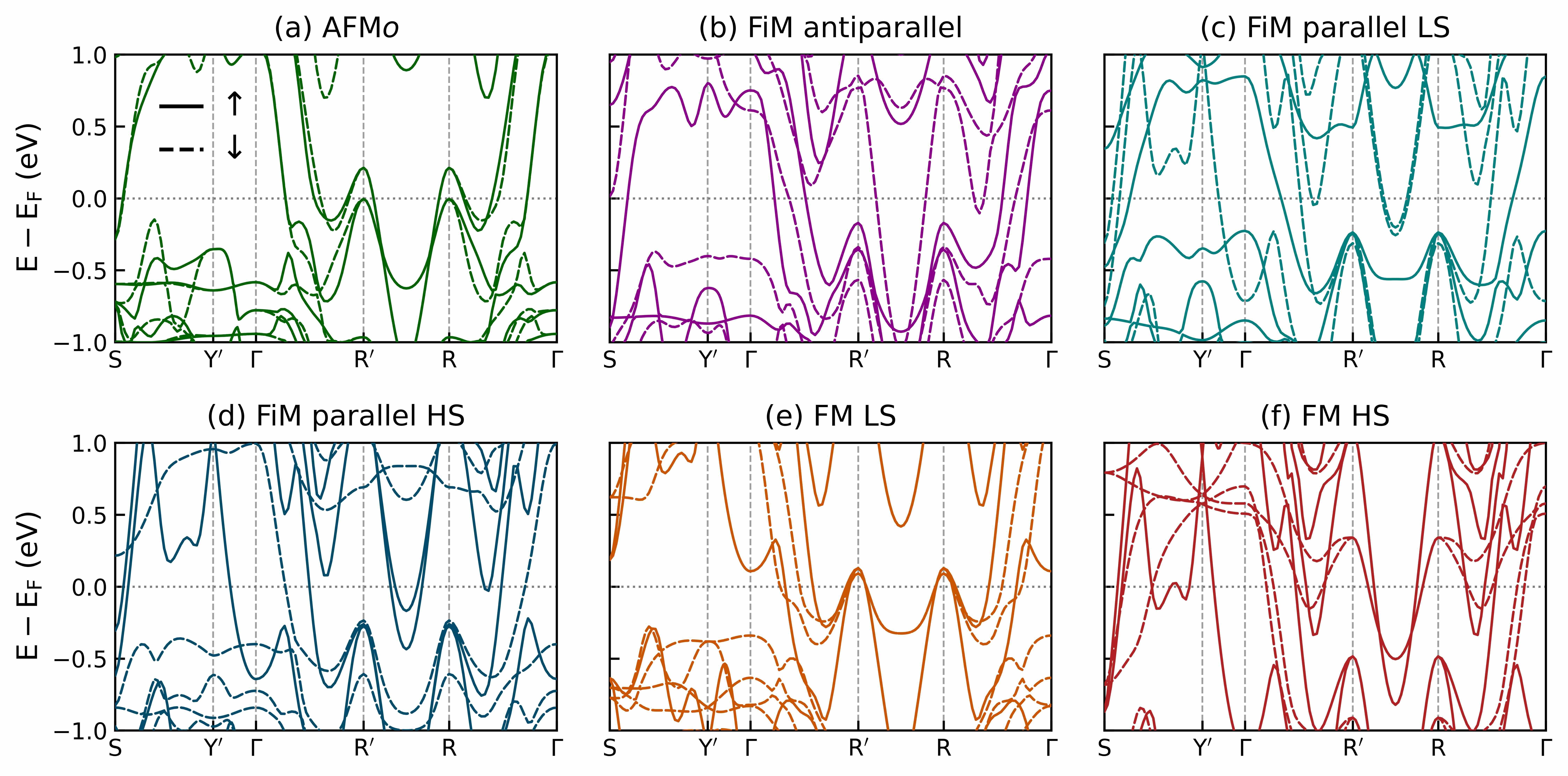}
\caption{\textbf{Spin-resolved VCA bands at $\bm{U=3}$~eV.} (a) \AFMo, (b) antiparallel FiM, (c,d) parallel FiM LS- and HS-like branches, and (e,f) FM LS- and HS-like branches for Co$_{0.15}$Fe$_{0.85}$Sb$_2$. Solid and dashed lines denote the two spin channels. The panels illustrate that similar total moments can arise from distinct occupation-matrix and electronic-structure solutions.}
\label{fig:VCA_bands_U3}
\end{figure*}

At the common value $U=3$~eV, all six branches in Fig.~\ref{fig:energy_vs_U} coexist as self-consistent solutions. Figure~\ref{fig:VCA_bands_U3} therefore compares their spin-resolved electronic structures at the same interaction strength, separating genuine differences between occupation-matrix minima from changes caused merely by using different $U$ values.

\begin{figure*}[t]
\centering
\includegraphics[width=0.96\textwidth]{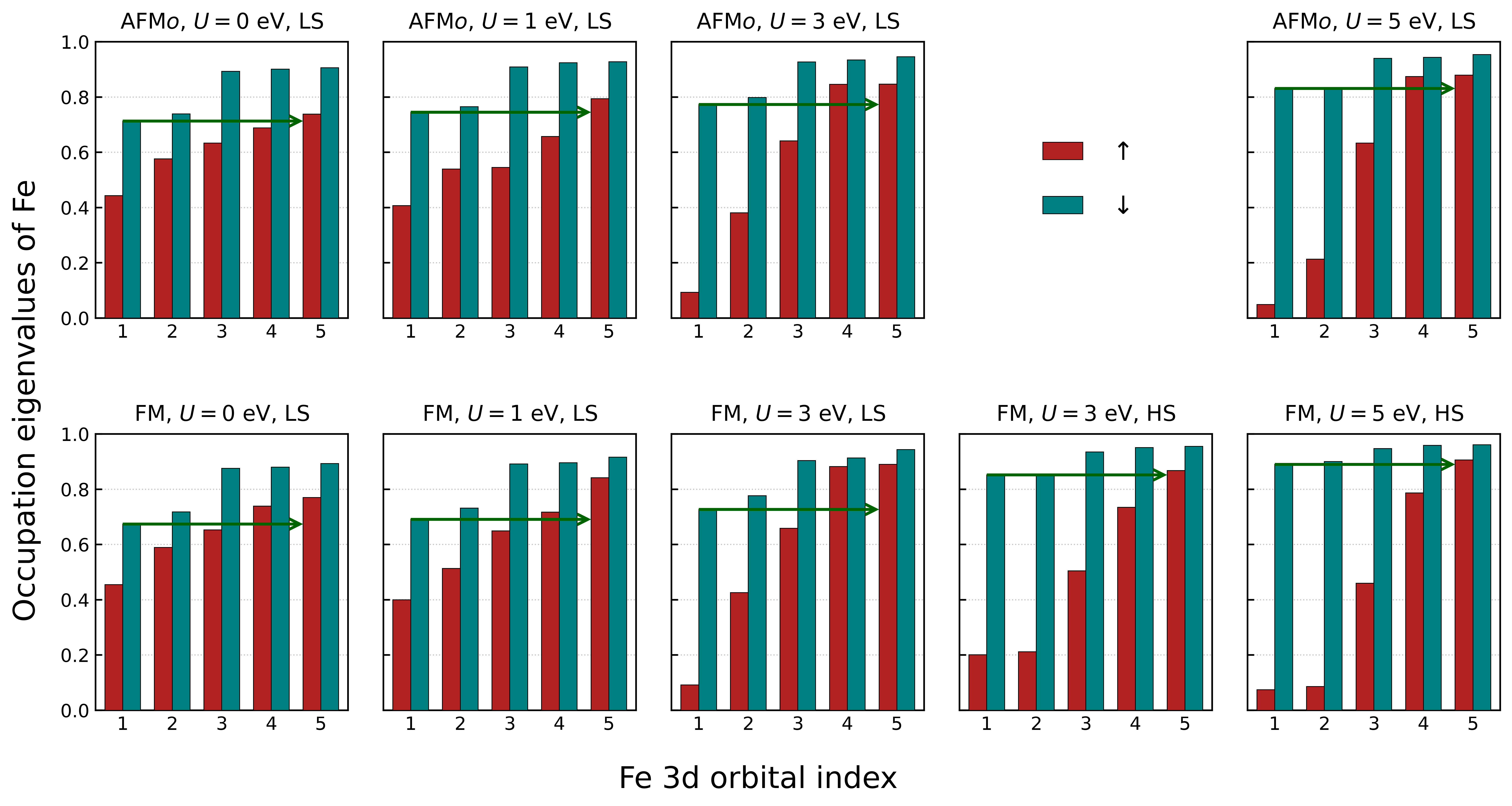}
\caption{\textbf{Fe-$\bm{3d}$ occupation eigenvalues.} Hubbard occupation-matrix eigenvalues for representative \AFMo\ and FM VCA branches as a function of $U$. The LS/HS notation is schematic for the metallic system; the arrow marks the minority/majority reordering used to identify the HS-like branch.}
\label{fig:Fe_occ}
\end{figure*}

\begin{figure*}[t]
\centering
\includegraphics[width=0.96\textwidth]{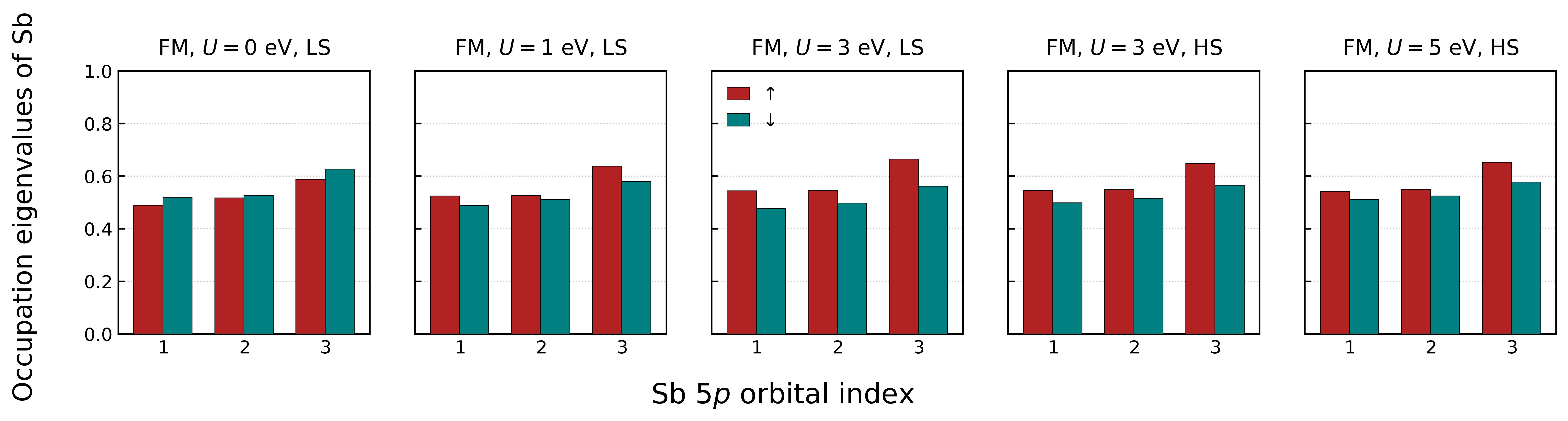}
\caption{\textbf{Sb-$\bm{5p}$ occupations.} Outer-$p$ occupation eigenvalues on Sb along the FM $U$ scan. Their weak variation, compared with Fig.~\ref{fig:Fe_occ}, locates the LS-to-HS-like redistribution predominantly on Fe.}
\label{fig:Sb_occ}
\end{figure*}

The Fe Hubbard occupation eigenvalues identify the LS-to-HS-like crossover [Fig.~\ref{fig:Fe_occ}]. On the HS-like branch the smallest minority-spin occupation becomes comparable to or larger than the most occupied majority-spin level. The Sb-$5p$ occupations remain nearly unchanged [Fig.~\ref{fig:Sb_occ}], showing that the crossover is primarily an Fe-$3d$ charge redistribution rather than a ligand-driven change.

\begin{figure}[t]
\centering
\includegraphics[width=\columnwidth]{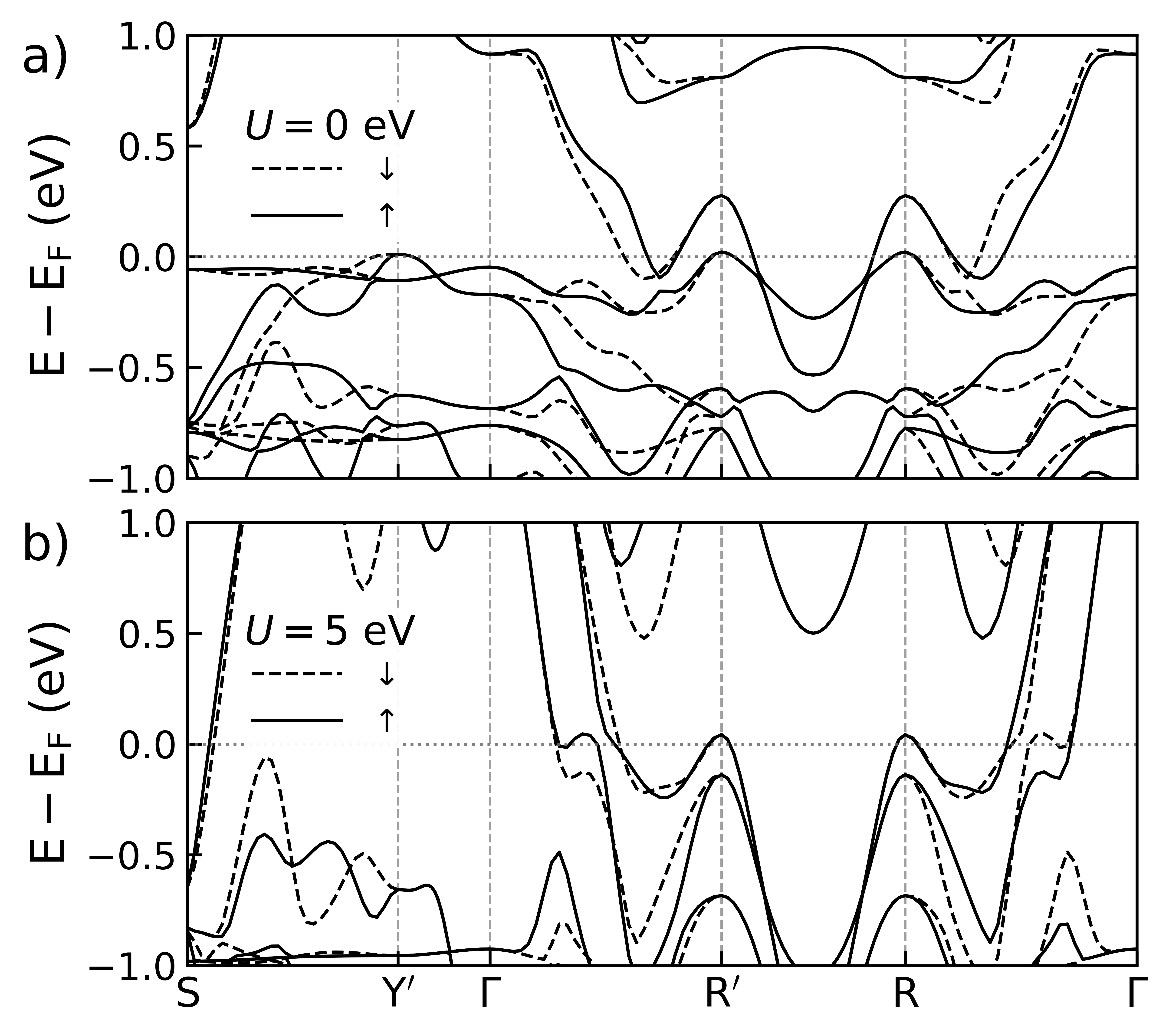}
\caption{\textbf{Co-doped AFM$\bm{o}$\ VCA bands.} Spin-resolved band structures calculated at (a) $U=0$ and (b) $U=5$~eV. The semilocal result is qualitatively consistent with Ref.~\cite{mazin2021prediction}, which considered a doping concentration of $x=0.2$; solid and dashed lines split along the symmetry-allowed $\mathrm S$--$\mathrm Y'$, $\Gamma$--$\mathrm R'$, and $\mathrm R$--$\Gamma$ segments.}
\label{fig:VCA_U0_U5_bands}
\end{figure}

The semilocal DFT+VCA \AFMo\ bands reproduce the characteristic symmetry-selective splitting reported in Ref.~\cite{mazin2021prediction}, despite the slightly different Co concentration, along $\mathrm S$--$\mathrm Y'$, $\Gamma$--$\mathrm R'$, and $\mathrm R$--$\Gamma$ [Fig.~\ref{fig:VCA_U0_U5_bands}]. Increasing $U$ renormalizes and reorders bands but preserves the allowed splitting paths.

\begin{figure*}[t]
\centering
\includegraphics[width=0.96\textwidth]{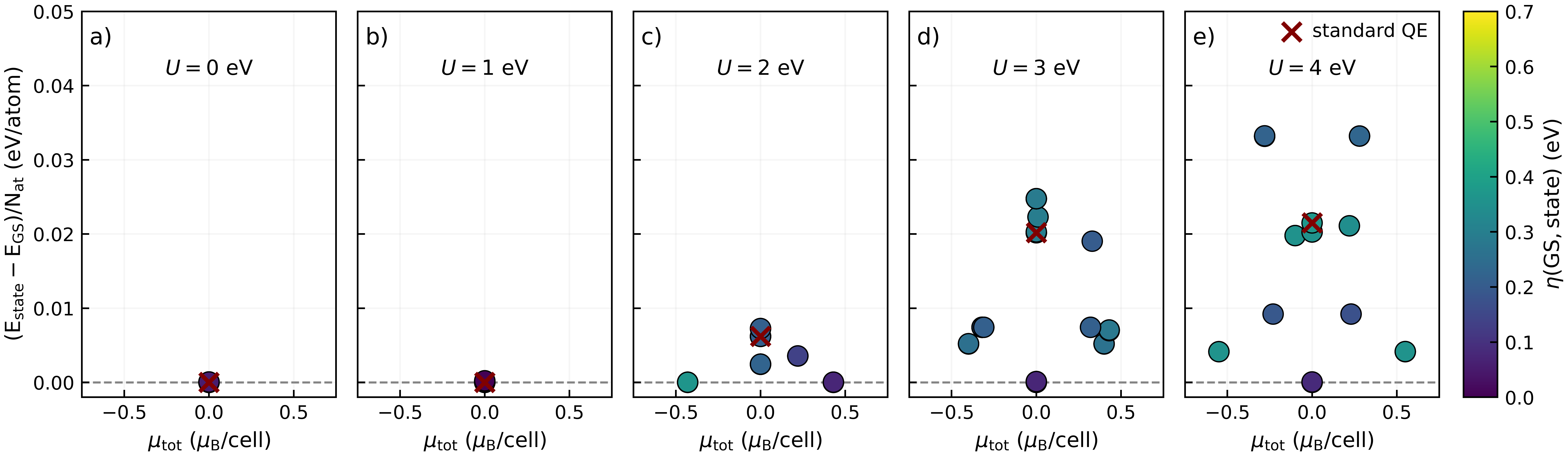}
\caption{\textbf{Co-doped VCA landscapes from $\bm{U=0}$ to $\bm{U=4}$~eV.} Each point is a distinct \texttt{Romeo} solution of the unconstrained energy functional, with the same axes and occupation-matrix-distance color convention as Fig.~\ref{fig:romeo}. Together with the $U=5$~eV main-text panel, the sequence demonstrates the nonmonotonic \AFMo/FiM\ ordering.}
\label{fig:Co_romeo_Uscan}
\end{figure*}

\begin{figure*}[t]
\centering
\includegraphics[width=0.96\textwidth]{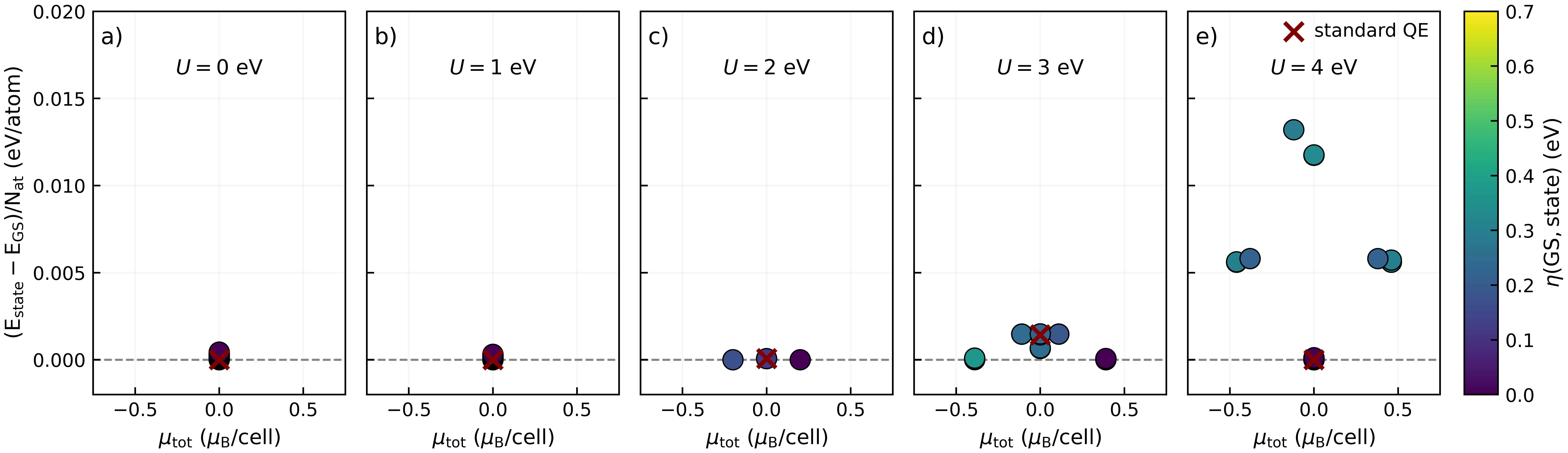}
\caption{\textbf{Cr-doped VCA landscapes from $\bm{U=0}$ to $\bm{U=4}$~eV.} Each point is a distinct \texttt{Romeo} solution of the unconstrained energy functional, with the same axes and occupation-matrix-distance color convention as Fig.~\ref{fig:romeo}. The VCA minimum is \AFMo\ at $U=0,1,4,$ and $5$~eV but FiM at the intermediate values $U=2$ and $3$~eV, confirming that the VCA magnetic classification is not robust against the interaction strength.}
\label{fig:Cr_romeo_Uscan}
\end{figure*}

\subsection{Robust occupation-matrix search}

The VCA replaces every Fe/dopant site by an averaged ionic potential with the target valence charge. It retains the six-atom primitive cell and uniform electron or hole doping, but removes local dopant environments. For each material and $U$, \texttt{Romeo} searches distinct initial Hubbard occupation matrices, releases the constraints, and converges them self-consistently \cite{ponet2024landscape}. Duplicate solutions are identified from energy, magnetization, local moments, and occupation-matrix distances. For solution $s$ we plot
\begin{equation}
 \Delta E_s=\frac{E_s-E_{\min}}{N_{\mathrm{at}}},
 \qquad
 M_s=\sum_I m_{I,s},
\end{equation}
and color the marker by the occupation-matrix distance $\eta(s,\mathrm{GS})$.

The ground-state identity changes nonmonotonically with $U$. For Co doping, \AFMo\ is lowest at $U=0,1,3,$ and $4$~eV, while FiM is lowest at $U=2$ and $5$~eV [Fig.~\ref{fig:Co_romeo_Uscan}]. For Cr doping, \AFMo\ is lowest at $U=0,1,4,$ and $5$~eV and FiM at $U=2$ and $3$~eV [Fig.~\ref{fig:Cr_romeo_Uscan}].

\section{Symmetry-first explicit-alloy supercells}
\label{sec:supercells}

\subsection{Minimal cell and exact altermagnetic constraint}

The primitive crystallographic cell contains Fe$_2$Sb$_4$. An index-20 supercell therefore has 40 transition-metal and 80 Sb sites; $x=0.15$ requires six dopants. Exact compensation assigns three dopants to each opposite-spin sublattice, but zero moment alone is not sufficient for altermagnetism. A single parent operation
\begin{equation}
 g=\{R|\bm\tau\},\qquad R\neq I,\quad R\neq-I,
 \label{eq:global_operation}
\end{equation}
must exchange the complete decorated sublattices,
\begin{equation}
 g:\{\mathrm{Fe}_{\uparrow},M_{\uparrow}\}
 \longleftrightarrow
 \{\mathrm{Fe}_{\downarrow},M_{\downarrow}\},
 \qquad g:\mathrm{Sb}\rightarrow\mathrm{Sb}.
 \label{eq:global_mapping_supp}
\end{equation}
For the $4\times1\times5$ construction, a representative operation in supercell fractional coordinates is
\begin{equation}
 (x,y,z)\mapsto
 \left(-x+\frac18,\ y+\frac12,\ -z+\frac1{10}\right)\bmod1.
\end{equation}
It is a twofold operation about $b$ combined with a fractional translation. Its two-site orbits accommodate exactly three up/down dopant pairs. A $2\times2\times5$ cell contains the same number of atoms but has incompatible orbit sizes for six dopants and cannot represent exact symmetry-defined \AFMo\ order at this composition. \texttt{AMCheck} independently verifies the $4\times1\times5$ construction \cite{amcheck1,amcheck2}.

\subsection{Enumeration and symmetry classification}

The generator reconstructs 20 up and 20 down metal sites from a labeled seed and considers
\begin{equation}
 N_{\mathrm{raw}}=\binom{20}{3}^{2}=1\,299\,600
\end{equation}
compensated chemical assignments. Parent operations are obtained with \texttt{spglib} \cite{togo2018spglib}; operations must map every up site to a down site and vice versa, preserve chemical identity and Sb, and have $R\neq\pm I$. We used a generation tolerance of $10^{-4}$~\AA\ and a Cartesian mapping tolerance of $10^{-3}$~\AA. The accepted decorations are quotiented by parent symmetries, supercell translations, permutations of identical atoms, and global spin reversal. This gives 26\,640 raw \AFMo-compatible decorations and 342 inequivalent spin-and-domain classes. All pass an independent \texttt{AMCheck} test. Grouping configurations by their complete spin-exchanging operation set yields 34 sets, dominated by a $C_{2b}$ screw-type family and a glide family normal to $a$. In fact, explicit substitution lowers the ordinary chemical space group from the parent $Pnnm$, because the surviving operations must separately preserve Fe, $M$, and Sb. The 342 classes are distributed as
\begin{equation*}
\begin{array}{lccc}
\hline
\text{Space group} & \text{No.} & \text{Crystal system} & \text{Classes} \\
\hline
P2_1   & 4  & \text{monoclinic}   & 276 \\
Pc     & 7  & \text{monoclinic}   & 48  \\
Pmn2_1 & 31 & \text{orthorhombic} & 18  \\
\hline
\end{array}
\end{equation*}
The $P2_1$ classes retain the screw-type operation, $Pc$ the glide, and $Pmn2_1$ both. These periodically ordered microscopic subgroups are compatible with an average experimental $Pnnm$ alloy structure because diffraction averages over substitutional disorder \cite{roy2026narrow}.

\subsection{Magnetic initializations and persistence of Fe moments}
\label{sec:magnetic_initializations}

Except for the independently optimized SQS cells (see Supplemental Sec.~\ref{sec:free_energy_supp}), auxiliary states were generated from the chemical decoration of the lowest-energy sampled \AFMo\ state. The tested initializations were: NM; FM with all Fe and $M$ moments parallel; FM(NM-Fe), with only the dopant magnetic; FM(NM-$M$), with only Fe magnetic; \AFMo(NM-Fe) and \AFMo(NM-$M$), retaining the inherited up/down pattern only on the magnetic species; and FiM(FM-Fe) or FiM(FM-$M$), obtained by ferromagnetically aligning one species while retaining the other species' \AFMo-derived pattern. The FiM-derived initializations test explicit-alloy analogues of the uncompensated state exposed by VCA-\texttt{Romeo}.

All converged NM, FM, FiM-derived, and SQS alternatives lie well above the low-energy \AFMo/\LDSC\ manifold. In particular, none of the tested initializations recover the VCA-like uncompensated FiM solution in which inequivalent Fe and $M$ moment magnitudes leave a small residual magnetization: FiM initializations either converge to a different magnetic class or remain high in energy. Most importantly, a dopant-only state analogous to Cr-doped RuO$_2$ is not stabilized: when Fe is initialized nonmagnetically, Fe moments either reappear during self-consistency or the calculation does not sustain a species-selective magnetic solution; conversely, suppressing the dopant moments does not produce a competitive Fe-only phase. Within the DFT+$U$ manifold, all low-energy explicit configurations retain finite Fe and dopant moments, demonstrating a substantially stronger intrinsic magnetic tendency of the Fe host than of Ru far from Cr-rich regions in RuO$_2$ and suggesting the weak FiM VCA minimum as an averaged-medium artifact.
To test this conclusion beyond the VCA and to check its sensitivity to species-dependent Hubbard parameters, Table~\ref{tab:explicit_U_comparison} compares the Co-doped explicit-alloy \AFMo, FM, and FiM(FM-Co) states. In the latter, all Co moments are aligned while the Fe network retains its antiparallel pattern. For both $(U_{\mathrm{Fe}},U_{\mathrm{Co}})=(3.5,3.5)$ and $(3.5,5.0)$~eV, \AFMo\ remains lowest; the FiM state is $0.15$--$0.22$~eV per 120-atom cell higher and the fully FM state is more than $1.4$~eV higher. This explicit-cell test confirms that ferromagnetic alignment is disfavored in Co$_{0.15}$Fe$_{0.85}$Sb$_{2}$.

\begin{table}[t]
\centering
\small
\caption{Co-doped explicit-alloy energies for different species-dependent Hubbard parameters. For each $(U_{\mathrm{Fe}},U_{\mathrm{Co}})$ pair, $\Delta E$ is referenced to the corresponding \AFMo\ state. FiM(FM-Co) has all Co moments parallel and the Fe moments in the inherited antiparallel pattern.}
\label{tab:explicit_U_comparison}
\resizebox{\columnwidth}{!}{
\begin{tabular}{lcccc}
\toprule
State & $U_{\mathrm{Fe}}$ (eV) & $U_{\mathrm{Co}}$ (eV) 
& $\Delta E$ (eV/cell) & $M_{\mathrm{tot}}$ ($\mub$/cell) \\
\midrule
\AFMo       & 3.5 & 3.5 & 0     & 0.00  \\
FiM(FM-Co)  & 3.5 & 3.5 & 0.155 & 1.07  \\
FM          & 3.5 & 3.5 & 1.623 & 79.67 \\
\addlinespace
\AFMo       & 3.5 & 5.0 & 0     & 0.00  \\
FiM(FM-Co)  & 3.5 & 5.0 & 0.215 & 4.04  \\
FM          & 3.5 & 5.0 & 1.452 & 80.72 \\
\bottomrule
\end{tabular}
}
\end{table}

\section{Construction and energetics of the LDSC bridge}
\label{sec:ldsc_supp}

\subsection{Shortest-swap search}

The fixed-spin-count space contains $\binom{20}{3}^2=1\,299\,600$ raw compensated decorations. Under exactly the same spatial and global-spin-reversal equivalence used for the \AFMo\ enumeration, these reduce to $8\,334$ inequivalent classes: 342 \AFMo\ and $7\,992$ fully compensated non-\AFMo\ configurations. The latter are approximately 23 times more numerous, so exhaustive DFT+$U$ relaxation is out of reach. The bridge construction instead samples the physically motivated subset of non-\AFMo\ states lying on minimum-swap paths between converged low-energy \AFMo\ endpoints.

A labeled configuration is represented by the three dopant sites on each fixed spin sublattice. The only elementary moves are same-sublattice chemical exchanges,
\begin{equation}
 M_{\uparrow}(i)\leftrightarrow\mathrm{Fe}_{\uparrow}(j),
 \qquad
 M_{\downarrow}(i)\leftrightarrow\mathrm{Fe}_{\downarrow}(j),
\end{equation}
so stoichiometry and the 20-up/20-down count remain fixed. For two \AFMo\ decorations $A$ and $B$, the exact minimum number of moves is
\begin{equation}
 d(A,B)=|C_{\uparrow}^{A}\setminus C_{\uparrow}^{B}|
       +|C_{\downarrow}^{A}\setminus C_{\downarrow}^{B}|,
\end{equation}
where $C_{\sigma}$ is the dopant-site set. The bridge algorithm selects converged low-energy \AFMo\ endpoints, enumerates every interior state satisfying $d(A,I)+d(I,B)=d(A,B)$ on their shortest swap paths, tests each against the complete parent operation set, and stops at the smallest endpoint distance containing a genuinely non-\AFMo\ interior. Thus $\{\text{bridge states}\}\subset\{7\,992\ \text{non-\AFMo\ classes}\}$, but the bridge search does not enumerate or relax that full set. The lowest-energy sampled interior is denoted \LDSC.

\subsection{Real-space motif, classification, and experimental meaning}

Main-text Fig.~\ref{fig:Co_landscape}(a--c) shows the local relation among the Co-doped minimum-energy \AFMo\ and \LDSC\ structures and a second \AFMo\ state. The first same-spin Fe--Co occupation swap converts the \AFMo\ minimum into the \LDSC\ minimum, and a second swap reaches the displayed \AFMo\ state at distance two. The corresponding Cr construction is analogous, with the energetic ordering reversed. These swaps leave most local antiparallel environments unchanged but remove and, after the second swap, restore a global spin-sublattice-exchanging operation.

Let $c_I$ be the chemical identity and $s_I=\pm1$ the spin label at metal site $I$. For a candidate spin-exchanging operation $g$, define the local overlap on region $\mathcal R$,
\begin{equation}
 Q_g(\mathcal R)=\frac{1}{N_{\mathcal R}}\sum_{I\in\mathcal R}
 \delta_{c_{gI},c_I}\frac{1-s_{gI}s_I}{2}.
 \label{eq:local_overlap}
\end{equation}
An exact \AFMo\ configuration has $Q_g=1$ for the full cell. In \LDSC, regions away from the swapped sites retain $Q_g\simeq1$, while regions containing the defect have $Q_g<1$ and no operation gives unity globally. Thus $Q_g$ diagnoses local \AFMo-symmetry overlap, not a magnetic correlation length. The repeated supercell is a periodic motif; a macroscopically short-ranged phase would require a nonperiodic ensemble of translated, symmetry-related, or multiple defects.

The Cr-doped experiment imposes four relevant constraints \cite{shawon2026evidence}: nearly compensated nonsaturating magnetization; a bulk magnetic volume fraction below $T_N\simeq3.5$~K; no coherent zero-field $\mu$SR precession but a broad quasistatic field distribution; and no magnetic Bragg intensity. \LDSC\ satisfies the corresponding static requirements---finite local Fe/Cr moments, negligible total moment, inequivalent local environments, and no global \AFMo\ operation---and is the lowest-energy sampled Cr configuration.

\section{Unfolded electronic structure}
\label{sec:unfolding_supp}

The 120-atom cell folds the primitive bands into a Brillouin zone 20 times smaller.

\noindent We unfold each Kohn--Sham state onto primitive momentum $\mathbf k$ with weight
\begin{equation}
 W_{n\mathbf K}(\mathbf k)=
 \sum_m\left|
 \left\langle\psi_{m\mathbf k}^{\mathrm{PC}}
 \middle|\Psi_{n\mathbf K}^{\mathrm{SC}}\right\rangle
 \right|^2.
\end{equation}
The calculation uses \texttt{BandUPpy} \cite{mondal2026banduppy}, which implements the BandUP formalism \cite{medeiros2014bandunfolding,medeiros2015spinor} and uses \texttt{IrRep} to read \texttt{Quantum ESPRESSO} wave functions \cite{iraola2022irrep}; the effective-band-structure interpretation follows Ref.~\cite{popescu2012unfolding}. Spectral maps use a Gaussian broadening of $0.01$~eV. To compare states whose alloy broadening prevents a one-to-one assignment of individual bands, we calculate the first moment of the unfolded weight in the fixed near-Fermi window $-0.5\le E-E_F\le0.5$~eV, using $0.03$~eV broadening on 601 energy points,
\begin{equation}
\mu_\sigma(k)=\frac{\int_{-0.5\,\mathrm{eV}}^{0.5\,\mathrm{eV}} E A_\sigma(k,E)\,dE}      {\int_{-0.5\,\mathrm{eV}}^{0.5\,\mathrm{eV}} A_\sigma(k,E)\,dE}.
\end{equation}
Here $A_\sigma(k,E)$ is the spin-resolved unfolded spectral function obtained by broadening the weights $W_{n\mathbf K}(\mathbf k)$. The centroid $\mu_\sigma(k)$ gives the energy center of the spin-$\sigma$ spectral weight in that window, while $|\Delta\mu(k)|=|\mu_\uparrow(k)-\mu_\downarrow(k)|$ compares the resulting spin splitting on the same footing in \AFMo\ and \LDSC. The same primitive path, Fermi-level convention, and spin projection are used for both states. An exact alternating relation links the two \AFMo\ spin spectra at symmetry-related momenta; no such relation constrains \LDSC.

\begin{figure}[t]
\centering
\includegraphics[width=\columnwidth]{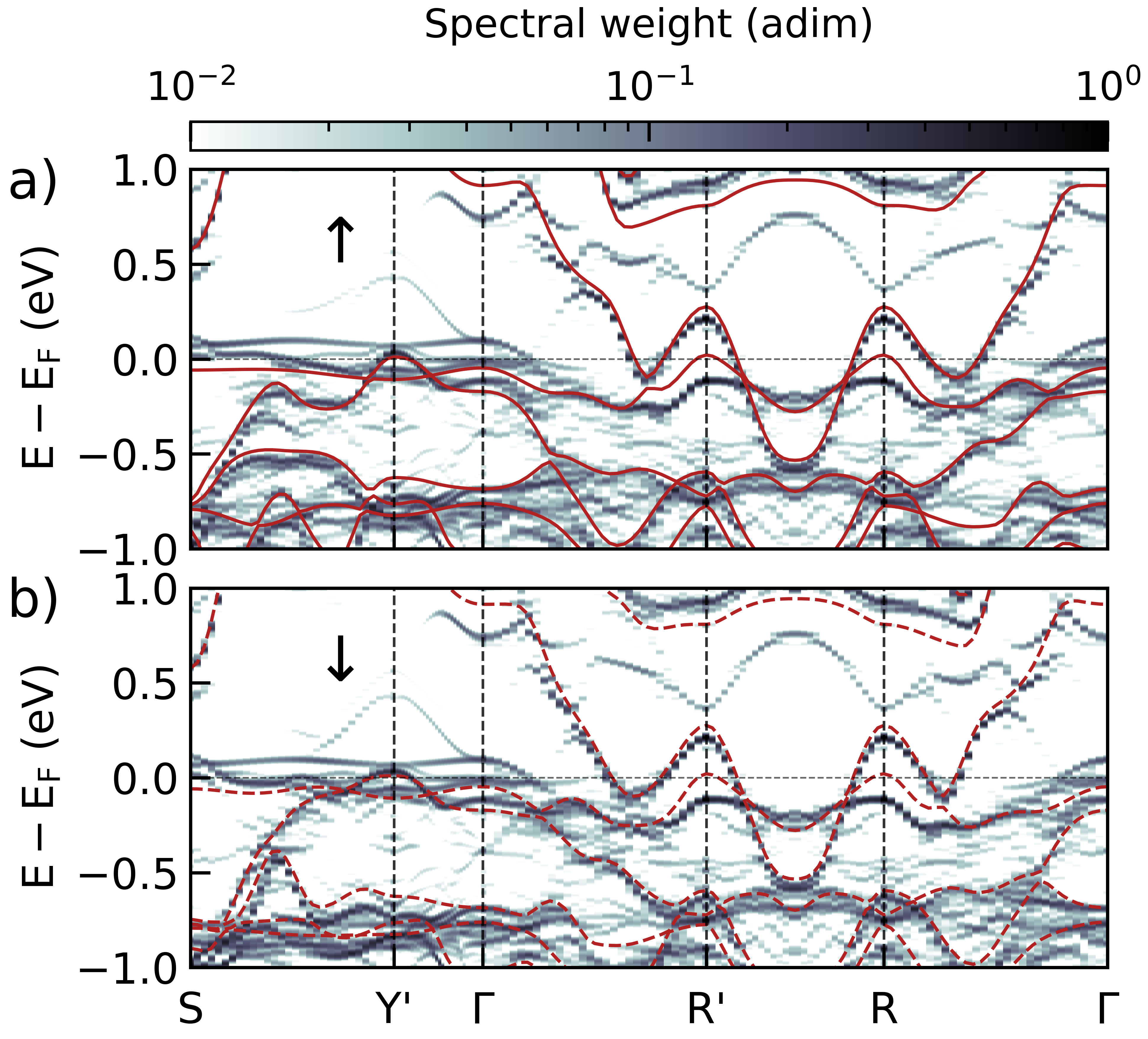}
\caption{\textbf{Co-doped semilocal-DFT unfolding.} Spin-up and spin-down unfolded spectral weights of the selected \AFMo\ supercell without $U$. The symmetry-allowed splitting occurs on the same primitive-cell path as in the VCA benchmark.}
\label{fig:Co_DFTonly}
\end{figure}

\begin{figure}[t]
\centering
\includegraphics[width=\columnwidth]{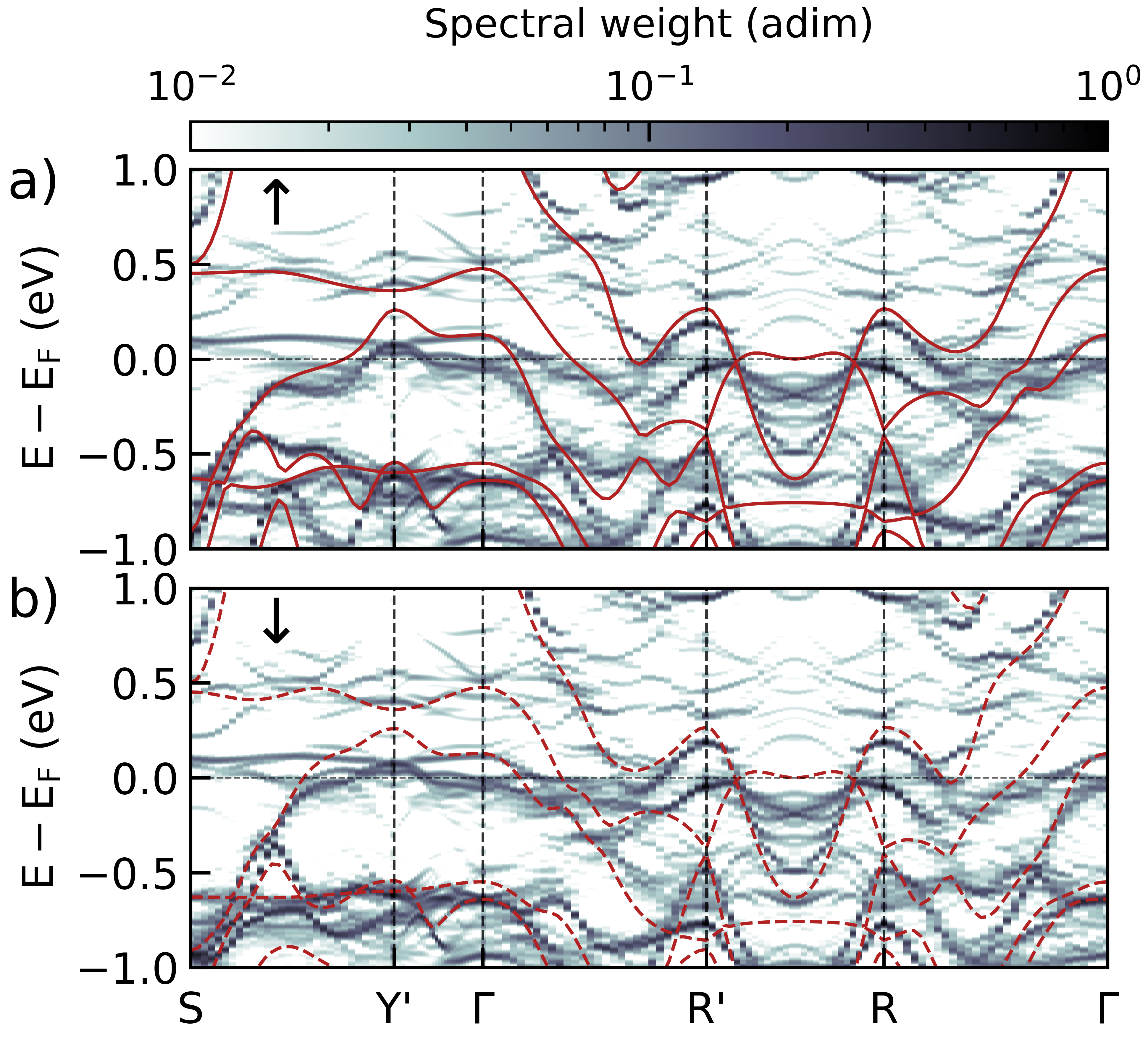}
\caption{\textbf{Cr-doped semilocal-DFT unfolding.} Spin-resolved unfolded weights of the selected \AFMo\ supercell without $U$, demonstrating the same symmetry-controlled momentum dependence before the Hubbard correction is applied.}
\label{fig:Cr_DFTonly}
\end{figure}

As a functional check, Figs.~\ref{fig:Co_DFTonly} and \ref{fig:Cr_DFTonly} show semilocal-DFT unfolded \AFMo\ spectra without the $U$ correction. They confirm that the symmetry-selected splitting paths do not originate from Hubbard-induced symmetry breaking, while the $U=5$~eV calculations sharpen the comparison with the magnetic energy landscape.

\section{Configurational free energies of the sampled ensemble and SQS references}
\label{sec:free_energy_supp}

\begin{figure}[t]
\centering
\includegraphics[width=\columnwidth]{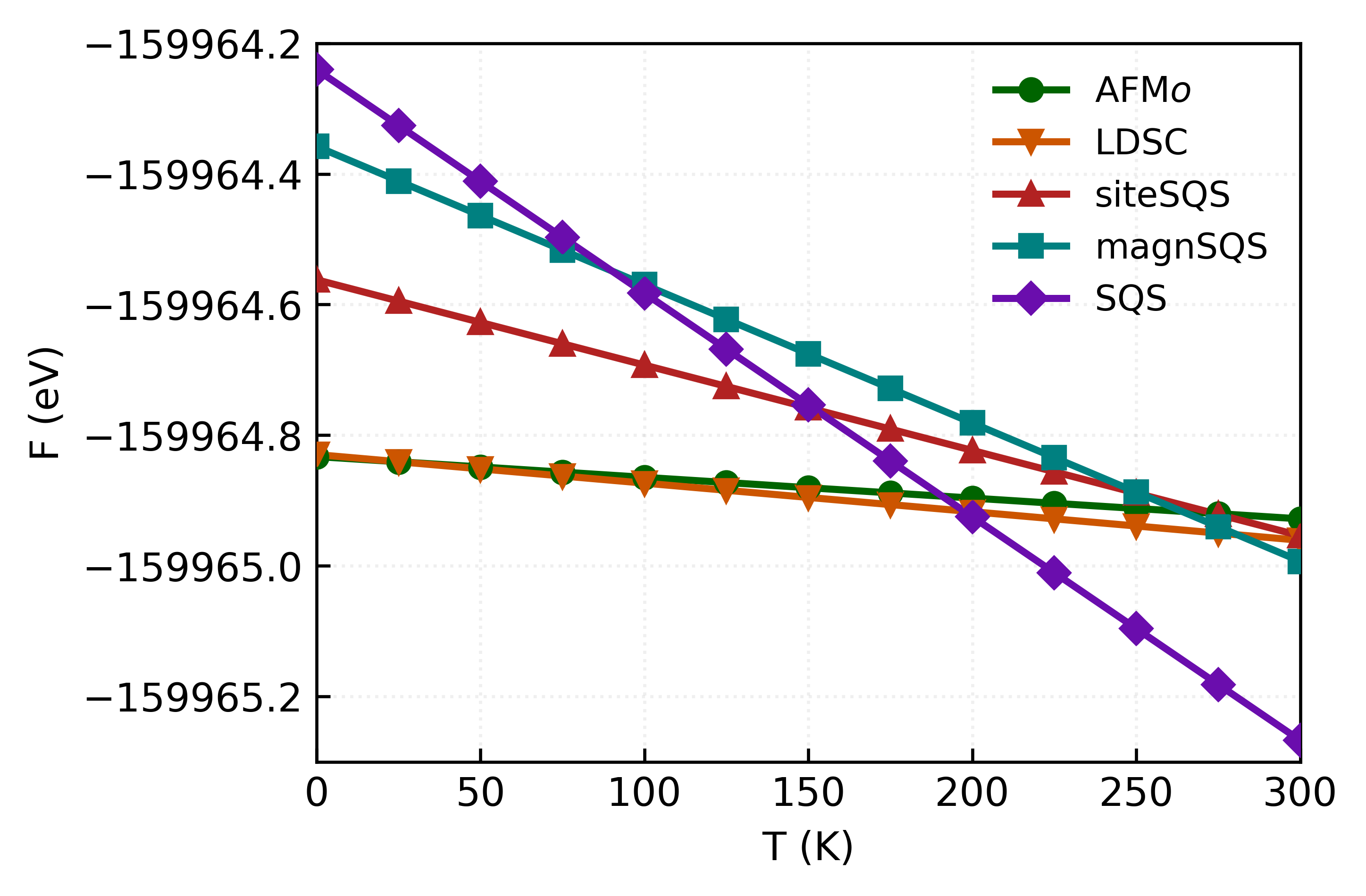}
\caption{\textbf{Configurational free energies for Co doping.} Free energies $A=E-TS_{\mathrm{conf}}$ of \AFMo\, \LDSC\, and the chemical, magnetic, and combined SQS reference macrostates. The larger \LDSC\ multiplicity produces the low crossing near $25$~K; the much higher SQS energies require substantially larger entropic terms before becoming competitive.}
\label{fig:free_energy_supp}
\end{figure}

Each class $i$ has DFT energy $E_i$ and multiplicity $\Omega_i$ under the same spin-and-domain equivalence used in the enumeration. For the finite set of classes considered, the partition function and class probability are
\begin{align}
 Z(T)&=\sum_i\Omega_i
 \exp\left[-\frac{E_i-E_{\min}}{\kb T}\right],\\
 P_i(T)&=\frac{\Omega_i
 \exp[-(E_i-E_{\min})/(\kb T)]}{Z(T)},
\end{align}
and the class free energy is
\begin{equation}
 A_i(T)=E_i-\kb T\ln\Omega_i.
 \label{eq:Ai_supp}
\end{equation}
The corresponding ensemble quantities are
\begin{align}
 F(T)&=E_{\min}-\kb T\ln Z(T),\\
 U(T)&=\sum_iP_iE_i,\\
 S_{\mathrm{conf}}(T)&=-\kb\sum_iP_i
 \ln\left(\frac{P_i}{\Omega_i}\right).
\end{align}
For chemical classes, $\Omega_i$ counts alloy realizations frozen during growth or annealing; it does not represent an entropy that can equilibrate through atomic diffusion at low temperature. The approximately $25$~K crossing between the Co-doped \AFMo\ ground-state class and the higher-multiplicity \LDSC\ class [Fig.~\ref{fig:free_energy_supp}] therefore does not predict a phase transition: the atomic occupations cannot rearrange between these configurations at such temperatures. It only indicates that, within the sampled set, the larger \LDSC\ multiplicity offsets its energy penalty. 

We also use three compensated SQS reference macrostates: (i) chemical SQS, with Fe/Co occupations disordered and the magnetic pattern fixed to that of the \AFMo\ ground state; (ii) magnetic SQS, with chemistry fixed to that of the \AFMo\ ground state and up/down moments disordered separately within each chemical species; and (iii) combined SQS, with both disordered magnetic moments and atomic sites. For Fe$_{34}$Co$_6$Sb$_{80}$, their ideal multiplicities are
\begin{align}
 \Omega_{\mathrm{chem}}&=\binom{40}{6},\\
 \Omega_{\mathrm{mag}}&=\binom{34}{17}\binom{6}{3},\\
 \Omega_{\mathrm{comb}}&=
 \frac{40!}{17!\,3!\,17!\,3!}
 =\Omega_{\mathrm{chem}}\Omega_{\mathrm{mag}}.
\end{align}
A literal individual SQS has multiplicity one. In Fig.~\ref{fig:free_energy_supp}, its DFT energy is instead used as a representative of the corresponding ideal-disorder macrostate, so the SQS curves are qualitative high-disorder references rather than quantitative equilibrium free energies.

\end{document}